# 730-nm optical parametric conversion from near- to short-wave infrared band

**J. M. Chavez Boggio,[1,*] J. R. Windmiller,[1] M. Knutzen,[2] R. Jiang,[1] C. Bres,[1] N. Alic,[1] B. Stossel,[3] K. Rottwitt,[2] and S. Radic[1]**

[1]*University of California, San Diego, 9500 Gilman Drive, La Jolla, CA 92093*
[2]*COM.DTU, Technical University of Denmark, 345V, DK-2800 Kgs. Lyngby, Denmark*
[3]*Lockheed Martin Corporation, IS&GS, Goodyear, AZ*
*Corresponding author:jmchavez@ece.ucsd.edu

**Abstract:** A record 730 nm parametric conversion in silica fiber from the near-infrared to the short-wave infrared band is reported and analyzed. A parametric gain in excess of 30 dB was measured for a signal at 1300 nm (with corresponding idler at 2030 nm). This conversion was performed in a travelling single-pass one-pump parametric architecture and high efficiency is achieved by a combination of high peak power and a nonlinear fiber with a reduced fourth-order dispersion coefficient.

## References and links

1. C. Weitkamp (Ed.), *Lidar: Range-Resolved Optical Remote Sensing of the Atmosphere* (Springer, 2005).
2. D. Stothard, M. Dunn, and C. Rae, "Hyperspectral imaging of gases with a continuous-wave pump-enhanced optical parametric oscillator," Opt. Express **12**, 947-955 (2004).
3. V.W.S. Chan, "Free-space optical communications," J. Lightwave Technol. **24**, 4750-4762 (2006).
4. C. Langrock and M.M. Fejer, "Fiber-feedback continuous-wave and synchronously-pumped singly-resonant ring optical parametric oscillators using reverse-proton-exchanged periodically-poled lithium niobate waveguides," Opt. Lett. **32**, 2263-2265 (2007).
5. K.K.Y. Wong, M.E. Marhic, G. Kalogerakis, and L.G. Kazovsky, "Fiber optical parametric amplifier and wavelength converter with record 360 nm gain bandwidth and 50 dB signal gain," presented at the Conference on Lasers & Electro-Optics, Baltimore, Md., 1-6 June 2003.
6. T. Torounidis and P.A. Andrekson, “Broadband single-pumped fiber-optic parametric amplifiers,” IEEE Photon. Technol. Lett. **19**, 650-652 (2007).
7. J.M. Chavez Boggio, M. Knutzen, C. Bres, N. Alic, J.R. Windmiller, B. Stossel, K. Rottwitt, S. Radic, "All fiber parametric conversion from near to short wave infrared band," presented at the European Conference on Optical Communications, PD, Berlin, Germany 16-20 September 2007.
8. R. Jiang, R. Saperstein, N. Alic, M. Nezhad, C. J. McKinstrie, J.E. Ford, Y. Fainman, S. Radic, "Parametric wavelength conversion from conventional near-infrared to visible band," IEEE Photon. Technol. Lett. **18**, 2445-2447 (2006).
9. R. Jiang, N. Alic, C. J. McKinstrie, S. Radic, “Two Pump Parametric Amplifier with 40dB of Equalized Continuous Gain over 50nms,” presented at Optical Fiber Communication Conference, (OFC/NFOEC 2007), Anaheim, March 2007.
10. J. M. Chavez Boggio, J. D. Marconi, S. R. Bickham, and H. L. Fragnito, "Spectrally flat and broadband double-pumped fiber optical parametric amplifiers," Opt. Express **15**, 5288-5309 (2007)
11. J.M. Chavez Boggio and H.L. Fragnito, "Simple four-wave mixing based method for measuring the ratio between the third- and fourth-order dispersion in optical fibers," J. Opt. Soc. Am. B **24**, 2046-2054 (2007).
12. K. J. Blow and D. Wood, “Theoretical description of transient stimulated Raman scattering in optical fibers,” IEEE J. Quantum Electron. **25**, 2665-2673 (1989).
13. E.A. Golovchenko, P.V. Mamyshev, A.N. Pilipetskii, E.M. Dianov, “Mutual influence of the parametric effects and stimulated Raman scattering in optical fibers,” IEEE J. Quantum Electron. **26**, 1815-1820 (1990).
14. F. Vanholsbeeck, P. Emplit, and S. Coen, " Complete experimental characterization of the influence of parametric four-wave mixing on stimulated Raman gain," Opt. Lett. **28**, 1960-1962 (2003).
15. A.S.Y. Hsieh, G.K.L. Wong, S.G. Murdoch, S. Coen, F. Vanholsbeeck, R. Leonhardt, and J.D. Harvey, "Combined effect of Raman and parametric gain on single-pump parametric amplifiers," Opt. Express **15**, 8104-8114 (2007).

16. R. H. Stolen, J. P. Gordon, W. J. Tomlinson, and H. A. Haus, "Raman response function of silica-core fibers," J. Opt. Soc. Am. B **6**, 1159-1167 (1989).
17. The complex part of the Raman third order nonlinear susceptibility was obtained from available data in the commercial VPI software, while the real part was obtained using Kramers-Kronig relations.
18. A.V. Shakhanov, K. M. Golant, A. N. Perov, S. D. Rumyantsev, A. G. Shebunyaev, I. I. Cheremisin, and S. A. Popov, "All-silica optical fibers with reduced losses beyond two microns," Proc. SPIE **1893**, 85-89 (1993).
19. J. D. Shephard, W. N. MacPherson, R. R. J. Maier, J. D. C. Jones and D. P. Hand, M. Mohebbi, A. K. George, P. J. Roberts and J. C. Knight, "Single-mode mid-IR guidance in a hollow-core photonic crystal fiber," Opt. Express **13**, 7139-7144 (2005).
20. M Hirano, T Nakanishi, T Okunko, M Onishi, "Selective FWM-based wavelength conversion realized by highly nonlinear fiber," Proc. European conference on optical communications (ECOC), Cannes Sept. 2006.
21. J. E. Sharping, M.A. Foster, A.L. Gaeta, J. Lasri, O. Lyngnes, and K. Vogel, "Octave-spanning, high-power microstructure-fiber-based optical parametric oscillators," Opt. Express **15**, 1474-1479 (2007).
22. G. K. L. Wong, S. G. Murdoch, R. Leonhardt, J. D. Harvey, and V. Marie, "High-conversion-efficiency widely-tunable all-fiber optical parametric oscillator," Opt. Express **15**, 2947-2952 (2007).

## 1. Introduction

The short-wave infrared (SWIR) band plays a unique role in sensing [1], active hyperspectral image [2], and free-space communications [3]. Unfortunately, available laser sources and receivers operating in the SWIR band have inferior performance when compared with those that exist in the near-infrared (NIR) [1]. This is disadvantageous for advanced spectroscopic or LIDAR systems requiring high-power, tunable, modulated or multiple-tone seeds within the SWIR band. Indeed, advanced LIDAR systems operating near 1550 nm use readily available wavelength division multiplexing blocks and densely positioned, narrow linewidth lasers, multiplexers and receivers. However, such advanced LIDAR systems do not possess the equivalent components in the SWIR band, thus imposing a significant limitation on the technology. While crystal-based devices can, in principle, be employed to convert from the NIR to SWIR band, they are also characterized by low efficiency, slow-tuning rate and limited spectral response [4]. In contrast, an all-fiber, single-pass optical parametric amplifier (OPA) offers broad bandwidth and efficient conversion limited only by escalating silica loss in the SWIR band and by the NIR-SWIR phase-matching requirement. The loss can be addressed by designing a short (~10m) fiber device with a high (>25dB) parametric gain (conversion efficiency). The challenge that emanates from maintaining the phase-matching between distant NIR and SWIR waves requires the use of silica fibers with specially tailored high order dispersion.

OPAs may be designed for operation with one or two pump waves. The former architecture is simpler and more convenient for broadband operation since it is less sensitive to random longitudinal fluctuations of the zero dispersion wavelength ($\lambda_0$). Previous results on wideband OPAs using the one-pump configuration demonstrated a signal gain in excess of 20 dB over a bandwidth of 360 nm using a pulsed pump [5] and 10 dB gain over nearly 100 nm using a continuous-wave (CW) pump [6]. A low spectral resolution measurement [7] has previously indicated the feasibility of NIR-SWIR conversion in an all-fiber device, albeit without conclusive proof of the spectral purity of the SWIR idler. In addition, a record wavelength conversion between the visible and the NIR spectral bands was achieved with a CW pump [8]. On the other hand, OPAs designed with two pumps have the advantage of a much uniform spectral response. Thus, the widest and most uniform (-3dB bandwidth) gain spectrum achieved was 40 dB over 50 nm in the CW regime [9] and 25 dB over 115 nm with pulsed pumps [10].

This paper analyzes and reports the performance of a single-pass optical parametric NIR-SWIR wavelength converter using a single section of a dispersion-engineered highly nonlinear fiber (HNLF) having very low fourth order dispersion. High-resolution spectral measurements have confirmed, for the first time, the ability to translate an arbitrary set of narrow tones within the NIR band up to 2030 nm with gain exceeding 30 dB.

## 2. Experimental setup and results

The one-pump OPA (1P-OPA) device is constructed and integrated into the experimental setup shown in Fig. 1. The pump source, ($\lambda_p$) was modulated using a 0.5ns-duration square pulse with a 1/300 duty cycle (and extinction ratio higher than 12 dB) and subsequently amplified by two cascaded Erbium doped fiber amplifiers (EDFAs). The noise from the EDFAs was rejected using band-pass filters (BPFs) having 1-nm bandwidth at -3dB. The gain medium was a 15m long HNLF section characterized by an effective area $A_{eff}$ of 11 $\mu m^2$ (measured at 1550 nm). The third order dispersion coefficient $\beta_3$ = 0.038 $ps^3$/km (dispersion slope $S_0$ = 0.027 ps/nm-km), was measured with a commercially available chromatic dispersion analyzer (Advantest Q7750). The zero dispersion wavelength and the fourth-order dispersion coefficient, with respective uncertainties, were measured using the recently reported method in [11], resulting in $\lambda_0$ = 1582.8 ± 0.3 nm and $\beta_4 = (1.4 \pm 0.4)\times10^{-5}$ $ps^4$/km, respectively. The HNLF was wound in a coil having a 0.5-m diameter in order to reduce radiative loss in the SWIR band. The maximum power that could be injected into the HNLF was 200 W without observing any pulse distortion. The pump peak power and pulse shape were monitored using a fast detector (p-I-n diode and transimpedance amplifier) and an oscilloscope. The pump power measurement uncertainty was estimated at 10% in our setup and is attributed to uncertainties in determining splice losses in the HNLF and the couplers. The parametric gain was characterized using a combination of a tunable (1410-1650 nm) and fixed (1300-1390 nm) CW lasers. An optical spectrum analyzer (OSA), operating at 0.1 nm resolution, was used to measure the parametric spectrum below 1700 nm, while the extended-wavelength idler spectrum was measured using a SWIR spectrometer (Spectral Products TDK-420) operating at 0.6-nm resolution and using a TE-cooled PbSe detector. Since the gain measurements were performed with the OSA, the duty cycle of 1/300 gives rise to a 24.8 dB correction.

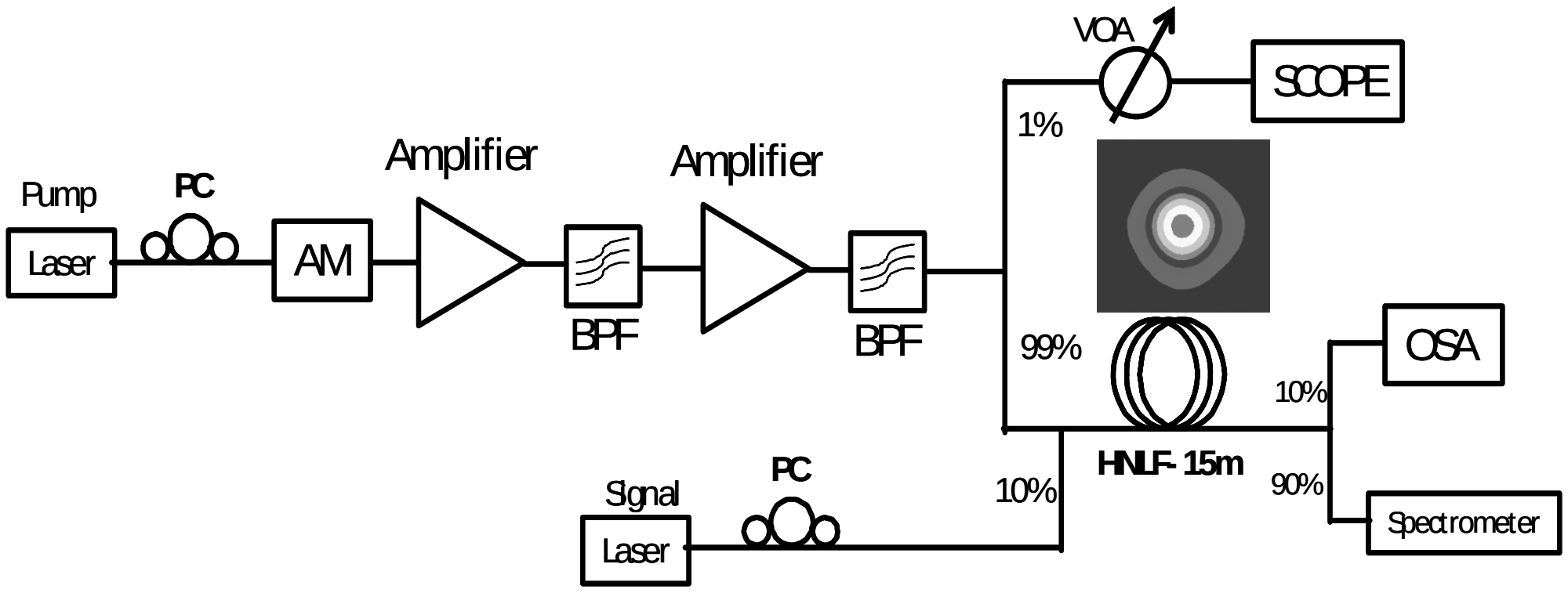

Fig. 1. Experimental setup. AM: amplitude modulator. BPF: band-pass filter. Inset: Far field mode profile for $\lambda_s$ = 1312 nm.

To maximize the conversion efficiency in the SWIR band, single-moded propagation and phase-matching between the interacting waves (pump, signal, and idler) are required. The inset of Fig. 1 illustrates the measured far-field mode profile that a signal at 1312 nm exhibits after propagating through the HNLF. The Gaussian pattern indicates that the cut-off wavelength was below 1300 nm, thus allowing for sufficient modal overlap between the lowest signal wavelength and the pump within the HNLF section. The conversion efficiency maximization is achieved by minimizing the signal-pump-idler mismatch for the four-photon mixing (FPM) process of

$$\kappa = \beta_2(\omega_p)(\omega_p - \omega_s)^2 + \beta_4(\omega_p)(\omega_p - \omega_s)^4/12 + 2\gamma P, \quad (1)$$

where $\beta_l$ ($l$ = 2,4) is the $l^{th}$ order dispersion coefficient. Exponential gain occurs when $-1 < \kappa/2\gamma P < 1$, while the maximum (matched) gain requires $\kappa = 0$. In general, the procedure for maximization of the 1P-OPA gain bandwidth consists of tuning the pump the closest possible to $\lambda_0$ (and this approach is only limited by the gain reduction effect due to the random $\lambda_0$ fluctuations in HNLF). However, while this procedure is valid for fibers possessing negative $\beta_4$, it does not apply for fibers with $\beta_4 > 0$, and especially so for the case when the nonlinear factor $\gamma P$ is large, as in our experiments. The latter is illustrated in Fig. 2 where $\kappa/2\gamma P$ is plotted, when the pump wavelength is varied within the anomalous dispersion region and in close vicinity (1583-1588 nm) of $\lambda_0$ (1583 nm). The simulation assumed the parameters describing the HNLF used in the experiments: $\beta_3$ = 0.04 ps$^3$/km, $\beta_4 = 1.4\times10^{-5}$ ps$^4$/km, and $\gamma P$ = 910 km$^{-1}$. In the case of $\beta_4 < 0$ (Fig. 2a), the maximum gain bandwidth is obtained when $\lambda_p = \lambda_0$, and as the pump is detuned away from $\lambda_0$ the bandwidth of the exponential gain decreases (noting that for any pump wavelength there are signal frequencies satisfying $\kappa = 0$). On the other hand, in the case that $\beta_4 > 0$ (Fig. 2b) the phase mismatch behavior is intrinsically different: the main characteristic is that complete phase matching ($\kappa = 0$) can not be satisfied for any signal wavelength if $\lambda_p < 1586$ nm. For example, locating the pump at $\lambda_p = \lambda_0$ implies that the exponential gain bandwidth is essentially reduced to zero because $\kappa/2\gamma P = 1 + \beta_4(\omega_p)(\omega_p - \omega_s)^4/24\gamma P > 1$. As $\lambda_p$ is tuned farther from $\lambda_0$, the $\beta_2(\omega_p)(\omega_p - \omega_s)^2$ and $\beta_4(\omega_p)(\omega_p - \omega_s)^4/24$ terms mutually compensate (i.e. for certain signal frequencies $0 < \kappa < 1$) resulting in an extension in the bandwidth and an increase in the peak gain. The optimal phase matching condition is achieved for a $\lambda_p$ slightly above 1586 nm and this pump wavelength gives rise to the broadest single-peaked bandwidth. For $\lambda_p$ = 1587 nm there are two signal-idler frequencies where phase-matching occurs and the gain spectrum will possess two pairs of peaks.

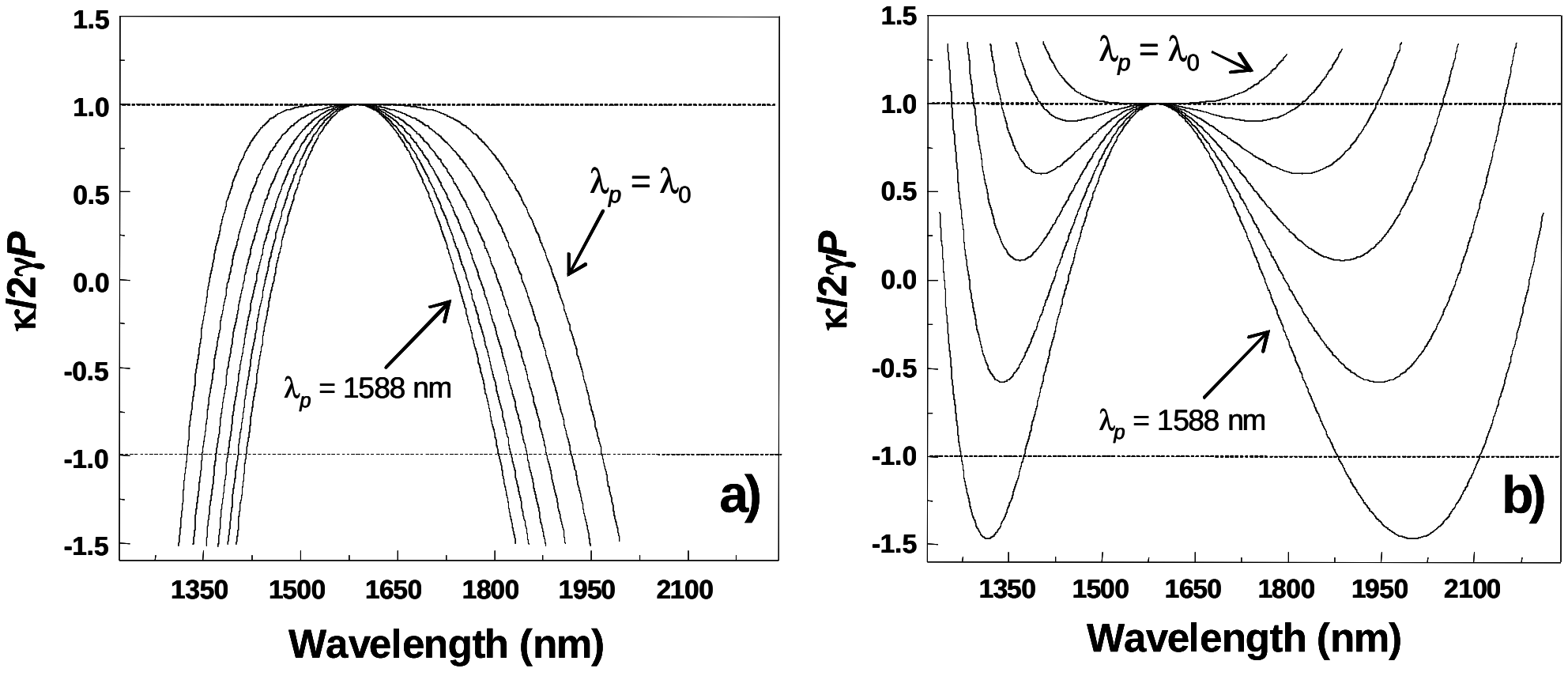

Fig. 2. $\kappa/2\gamma P$ as a function of $\lambda_s$ for $\lambda_p$ ranging from 1583 to 1588 nm. (a) $\beta_4 < 0$ and (b) $\beta_4 > 0$.

In the experiment, the parametric bandwidth was optimized by pump wavelength excursions around $\lambda_0$ and by measuring the consequent gain of CW signals centered at $\lambda_s$ = 1300 nm and at $\lambda_s$ = 1312 nm. The CW signal power at the input of the HNLF was –22 dBm. The maximum signal gain at those wavelengths was obtained for a pump positioned at $\lambda_p$ = 1585.45 nm and having a power of $P \cong 80$ W. Figure 3(a) illustrates the pump-on and -off spectra for the Fabry-Perot laser having multiple modes at 1300 nm, with an observed gain of 32.5 dB at 1302.6 nm. Figure 3(b) shows the pump-on and -off spectra for the single-mode DFB laser at 1312.6 nm, with an observed gain of 44.5 dB.

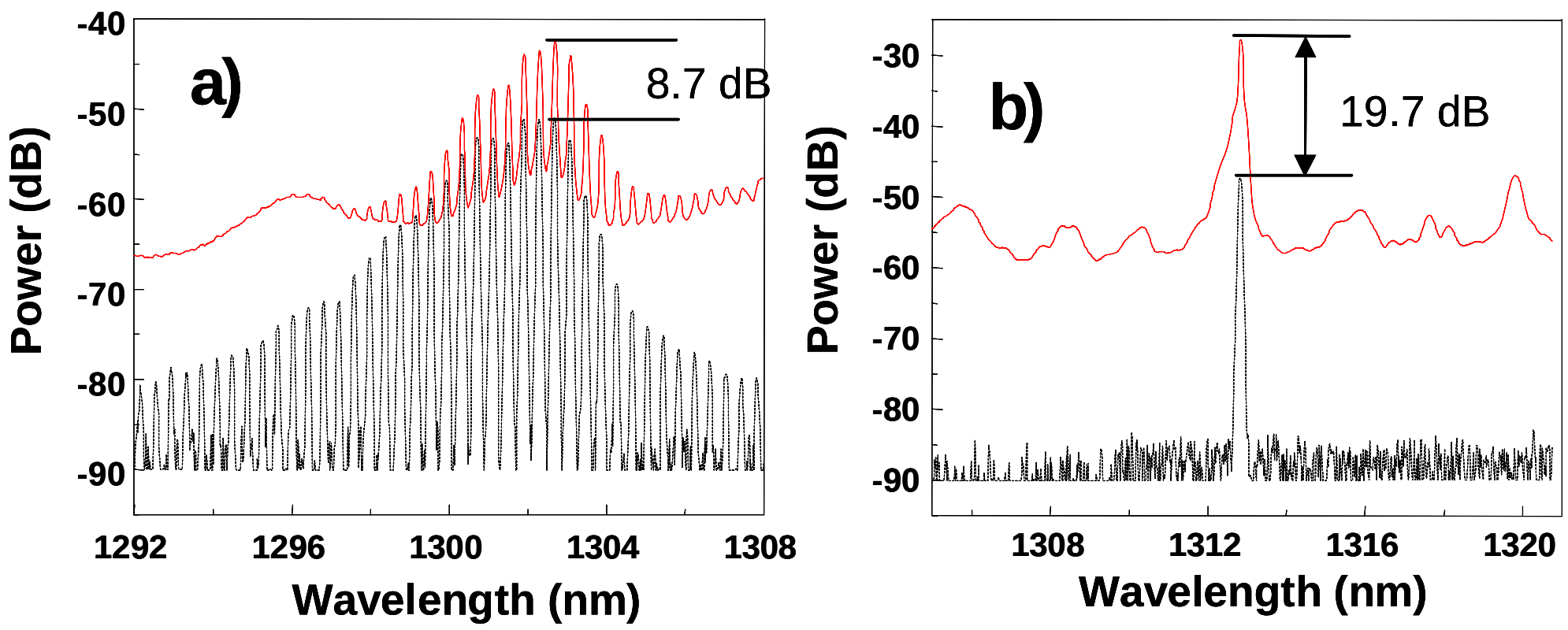

Fig. 3. Parametric amplification of (a) 1300 nm multiple-tone laser, (b) 1312 nm single-tone laser. The dashed black lines correspond to the pump off state.

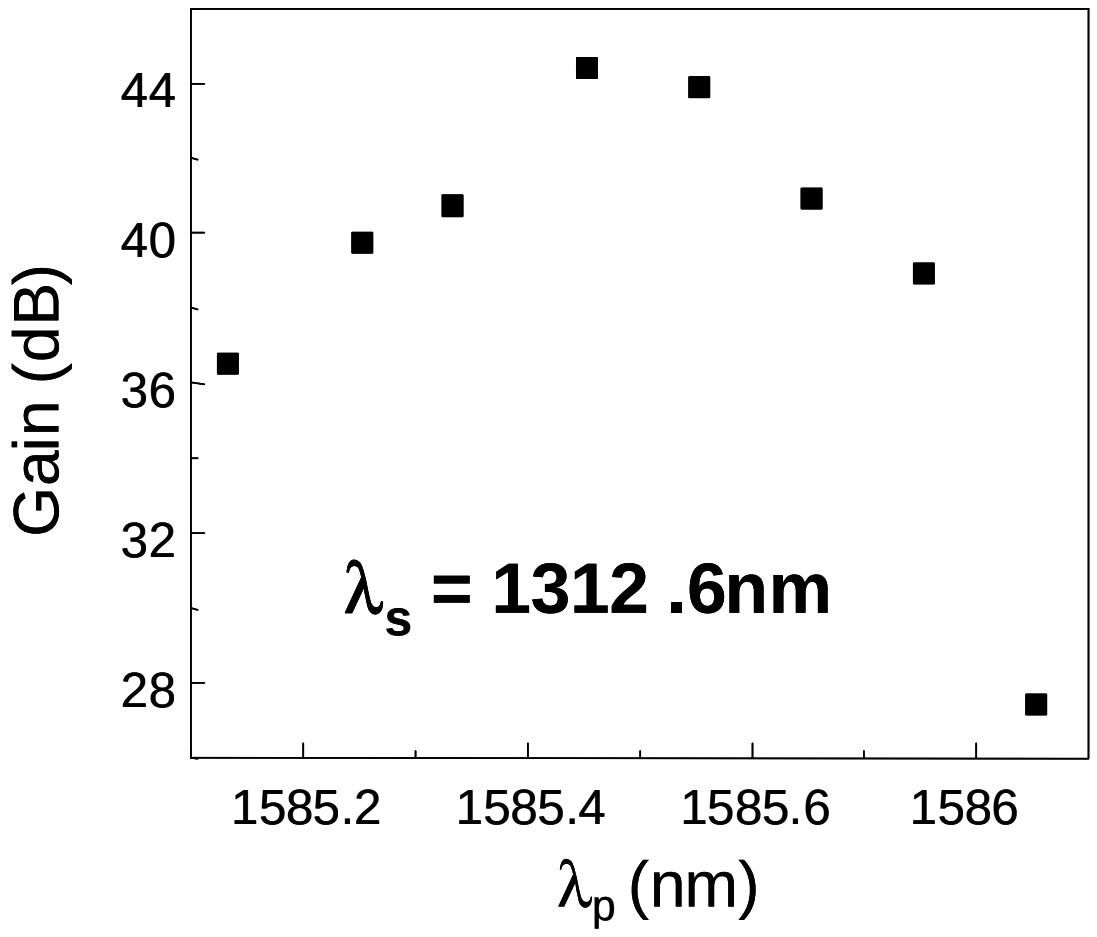

Fig. 4. Gain at $\lambda_s$ = 1312.6 nm as a function of $\lambda_p$. The pump power is maintained at 80 W with an error of less than 10%.

Figure 4 summarizes the signal gain measured at 1312.6 nm as a function of the pump wavelength. As noted by earlier qualitative prediction, as $\lambda_p$ is tuned away from $\lambda_0$, the gain increases with peak for $\lambda_p$ = 1585.45 nm and then decreases. The same measurement was performed for the signal at 1302 nm and an almost identical result was obtained: the highest gain was obtained for a pump wavelength at $\lambda_p$ = 1585.45 nm. These measurements allow us to find the optimum pump wavelength that yields the most efficient parametric conversion to the 2 microns band (for these two signal wavelengths).

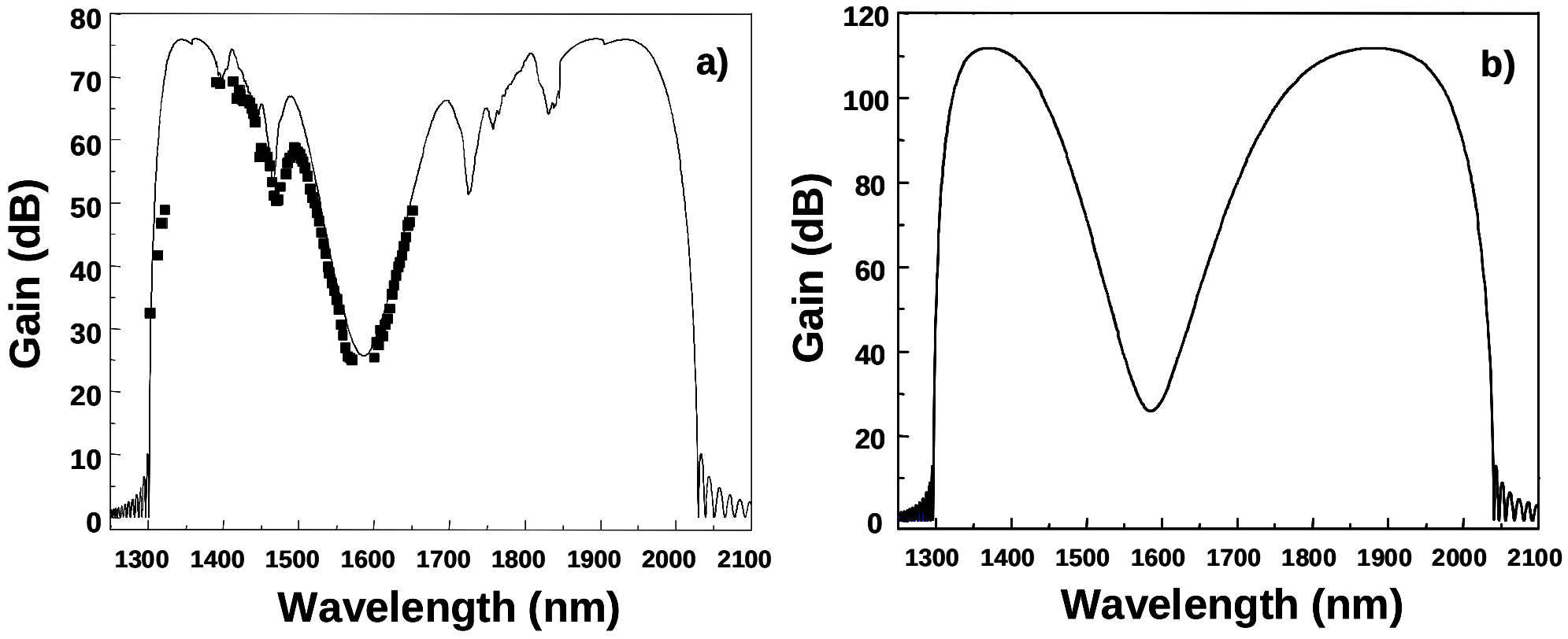

Fig. 5. (a) 1P-OPA gain spectrum: squares (experiment) and continuous lines (numerical fitting including the Raman third order nonlinear susceptibility). (b) 1P-OPA gain spectrum: Gain measurement error is < 1dB.

Figure 5(a) illustrates the signal gain spectrum measured using a combination of fixed and tunable lasers covering the spectral region from 1300 to 1650 nm (square marks) together with the calculated prediction including the Raman effect (continuous line) for the optimum pump wavelength $\lambda_p$ = 1585.45 nm. Pump and signal polarizations were aligned to maximize the parametric gain in all measurements. In order to avoid saturation of the parametric gain, in addition to preserving a minimum OSNR of 10 dB, the signal input power was adjusted between -15 and -55 dBm as the signal was tuned. Note in Fig. 5(a) that in addition to the characteristic modulation instability shape, the gain spectrum exhibited distortions due to a Raman-induced phase mismatch [12-16]. These distortions are more apparent within the 1400-1500 nm band, where a 10 dB gain dip is noted at 1470 nm. The theoretical prediction (shown with a solid line) was calculated using $\lambda_0$ = 1582.65 nm, a nonlinear coefficient $\gamma$ of 11.5 $W^{-1}km^{-1}$, and a fractional Raman contribution of $f$ = 0.25 to the total nonlinear susceptibility as fitting parameters [15]. The agreement with the experimental data is very good. More specifically, the calculation was aided by the exact Raman nonlinear susceptibility curve shown in the Fig. 6. The curve was obtained from measurements in a conventional fiber [17], mediating the disagreement between the initial calculations and the measurement in the 1490 nm spectral vicinity. As a comparison, Fig. 5(b) shows the gain spectrum when $f$ = 0: the main difference, when compared with Fig. 5(a), is that the peak gain is 35 dB higher.

In Fig. 5(a) the shortest measured signal wavelength was 1302.6 nm and exhibited 32.5 dB gain, with a corresponding converted wave at 2026 nm. Since the silica loss at 2.1 μm is estimated to be near 0.1 dB/m [18,19], the conversion efficiency at 2026 nm was greater than 30 dB. This qualifies, to the best to our knowledge, as a record in amplification bandwidth (730 nm) and wavelength conversion efficiency (> 30dB) of an all-fiber optical parametric amplifier using single-pass (travelling) architecture. We stress that this 730 nm bandwidth was limited by the fact that the available laser wavelengths were above 1300 nm. Since it can be deduced from Fig. 2(b) that gain for signals below 1300 nm should be appreciable, therefore measuring the gain at those shorter signal wavelengths should produce an OPA bandwidth larger than the 730 nm measured in Fig. 5(a).

To analyze the impact that $\lambda_0$ fluctuations can have on OPA bandwidth, we measured the gain spectrum using another 15 m segment of HNLF from the same spool of fiber, cut 100 m away relative to the previous 15 m segment. The gain bandwidth from the latter segment was almost identical. In particular, a gain of 42 dB at 1312 nm was observed.

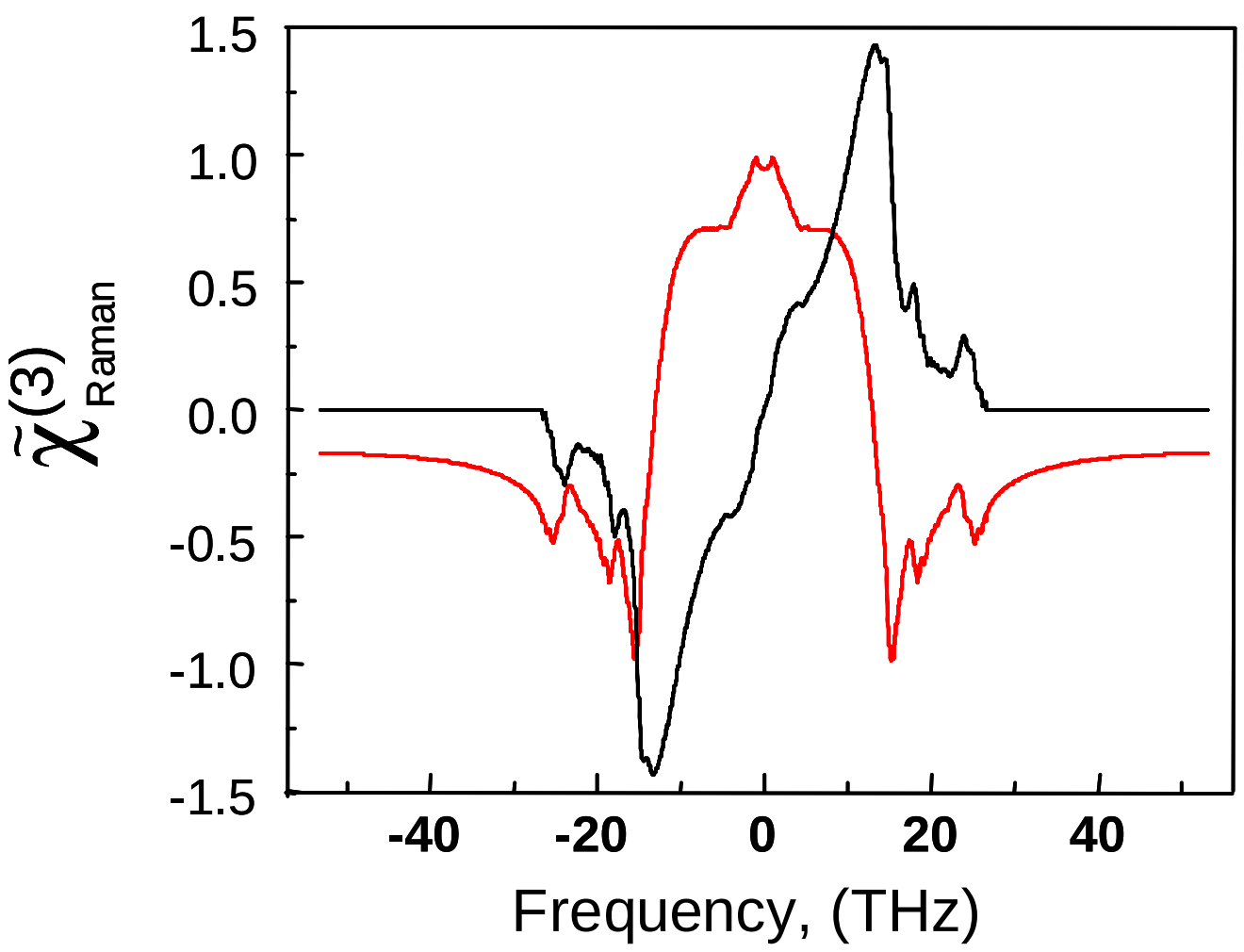

Fig. 6. Raman third order nonlinear susceptibility, $\chi^{(3)}_{Raman}$, as a function of detuning frequency: real component (red line), imaginary component (black line).

Figures 7(a) and 7(b) show the idler spectrum generated at $\lambda_i = \lambda_p\lambda_s/(2\lambda_s - \lambda_p)$ = 1798 nm (corresponding to the signal at $\lambda_s$ = 1417.3 nm and with –35 dBm input power) and $\lambda_i$ = 1999 nm (corresponding to $\lambda_s$ = 1312.6 nm and –15 dBm input power), respectively. The high coupling loss into the spectrometer and the dark current generated in the PbSe detector imposed a limit on the optical signal to noise ratio (OSNR) for the SWIR idler measurements. A lower OSNR is observed for the idler at 2000 nm when compared with idler OSNR at 1800 nm. This can be attributed to the fact that the output idler powers were different (0.8W versus 4W, respectively) due to the lower gain experienced at 2000 nm.

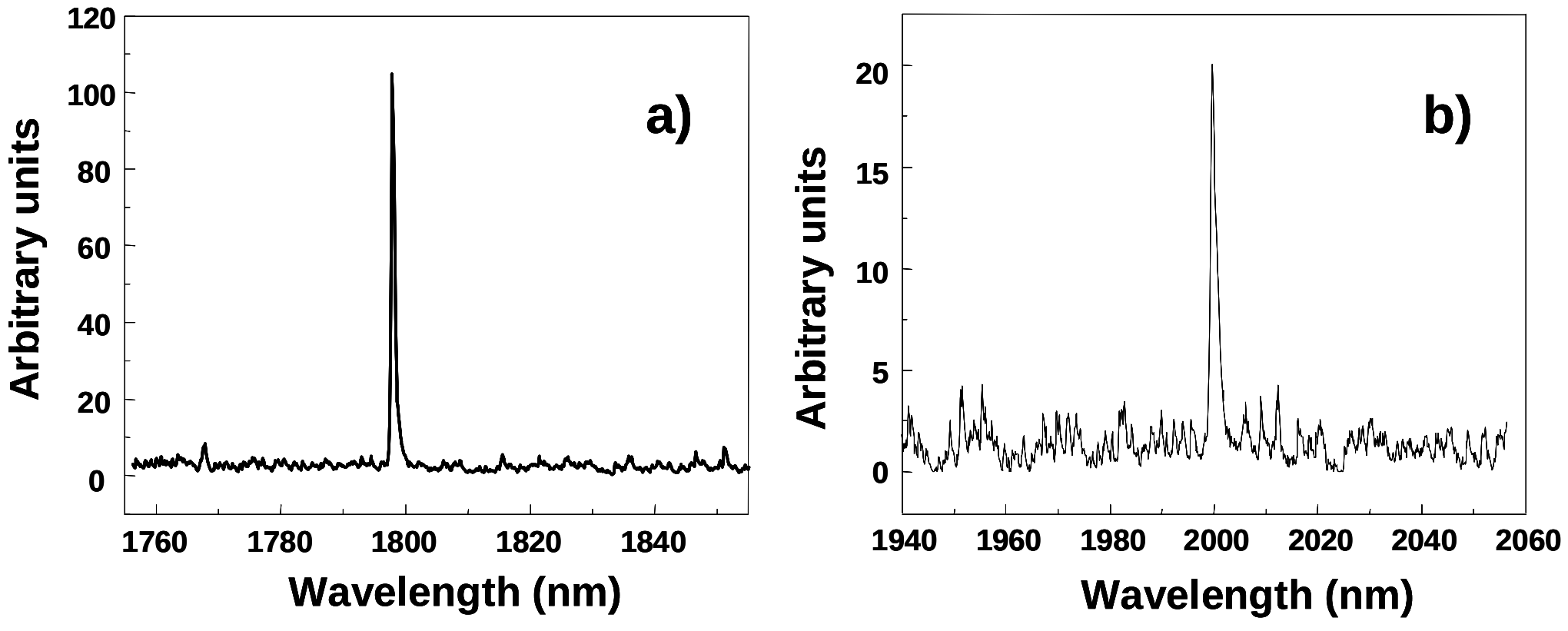

Fig. 7. Spectra of the converted waves into the SWIR band for (a) $\lambda_s$ = 1417 nm, (b) $\lambda_s$ = 1312.6 nm.

## 4. Discussion

It is well known that the OPA bandwidth scales as $\sqrt[4]{\gamma P/\beta_4}$ , therefore a further bandwidth extension may be obtained by increasing the pump power (and decreasing the fiber length). However, preliminary gain measurements with $L$ = 9 and $L$ = 12 m segments from the same fiber coil suggest that such a bandwidth improvement cannot be obtained, despite the fact that the pump power was increased to 150 W. For any pump wavelength, the signal gain at 1312 nm was almost zero. This may be attributed to amplification of background noise that resulted in an effective compression of the signal gain. The noise-mediated gain compression was observed whenever the peak gain exceeded ~70 dB, achieved with a pump power in excess of ~85 W when $\lambda_p$ = 1585.5 nm. Similar behaviour was observed for longer pump wavelengths at lower peak powers.

This is illustrated in Fig. 8(a) for the case of the segment with $L$ = 15 m, where the gain is plotted as a function of the pump power (from 34 to 90 W) when the pump is located at $\lambda_p$ = 1585.5 nm. As the pump power is increased, the parametrically amplified noise increases, however the noise in the spectral proximity of the pump increases at a more rapid rate. An indication of the pump spectral characteristics at the input of the HNLF is shown in Fig. 8(b). Note that there are side-lobes around the pump that resemble the well-known modulation instability (MI) phenomenon (and are generated within the booster amplifier). Upon closer inspection it can be observed that, besides the MI, there exists a noise pedestal around the pump which imposes an optical-pump-to-noise-ratio of around 20 dB. This input noise in the spectral vicinity of the pump should be the origin of the strong noise build-up observed in Fig. 8(a).

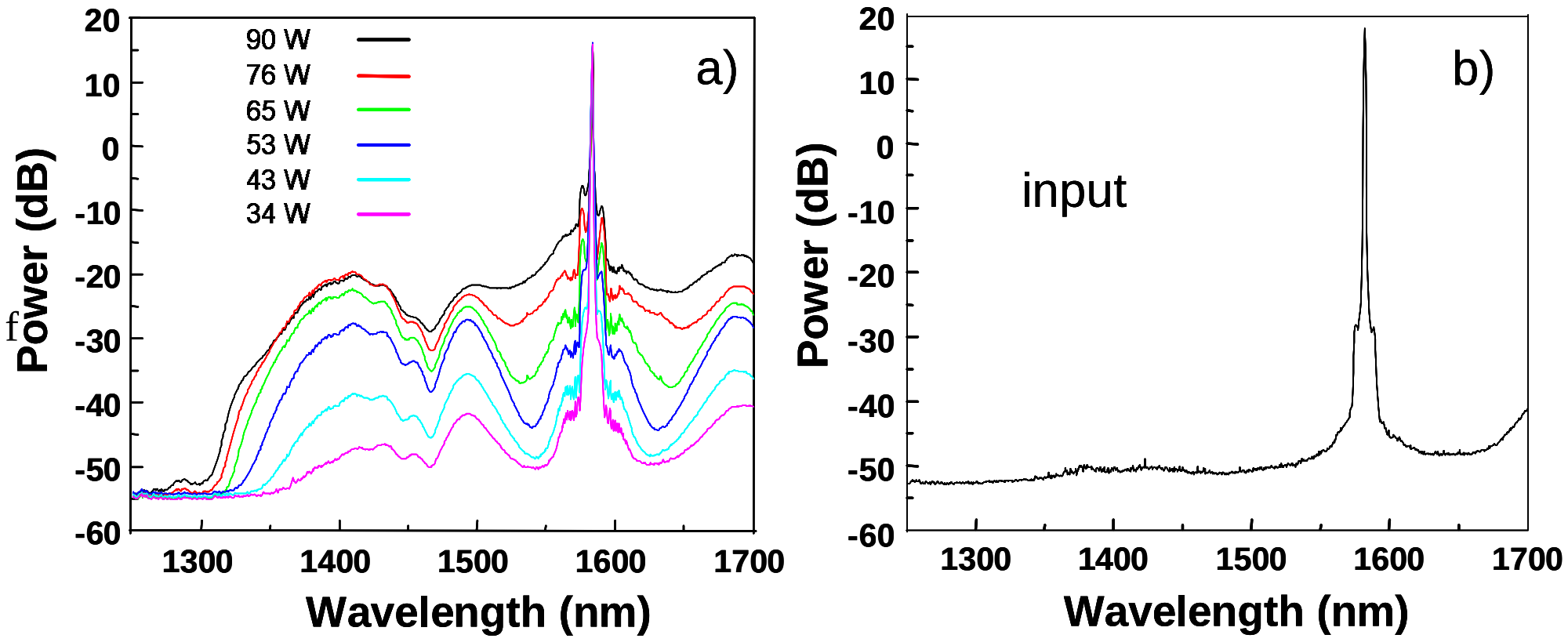

Fig. 8. (a) OPA gain spectrum as a function of the pump power for $\lambda_p$ = 1585.5 nm. The relative values of the pump power was measured with error less than 5%, while the absolute peak value (90 W) was measured with an error of ±10%. (b) Spectrum at the input of the HNLF when $P$ = 90 W.

Moreover, Fig. 9 illustrates the evolution of the noise spectrum for the case when the pump power was maintained at 80 W and the pump wavelength was tuned from 1584.4 to 1588.3 nm. As the pump is detuned to longer wavelengths the gain increases because the FWM process becomes increasingly phase-matched (see Fig. 2b), and some gain saturation starts to occur for this pump power when the pump wavelength is positioned at 1586 nm. Note that for some pump wavelengths there is noise gain in the 1270-1300 nm region corresponding to 2030-2110 nm in the opposite end of the spectrum. This noise amplification is more intense for the pump wavelengths at 1586-1587 nm. This corresponds to an 840-nm bandwidth of parametric noise amplification and should give an indication of the largest signal bandwidth that can be obtained with this 15m long HNLF.

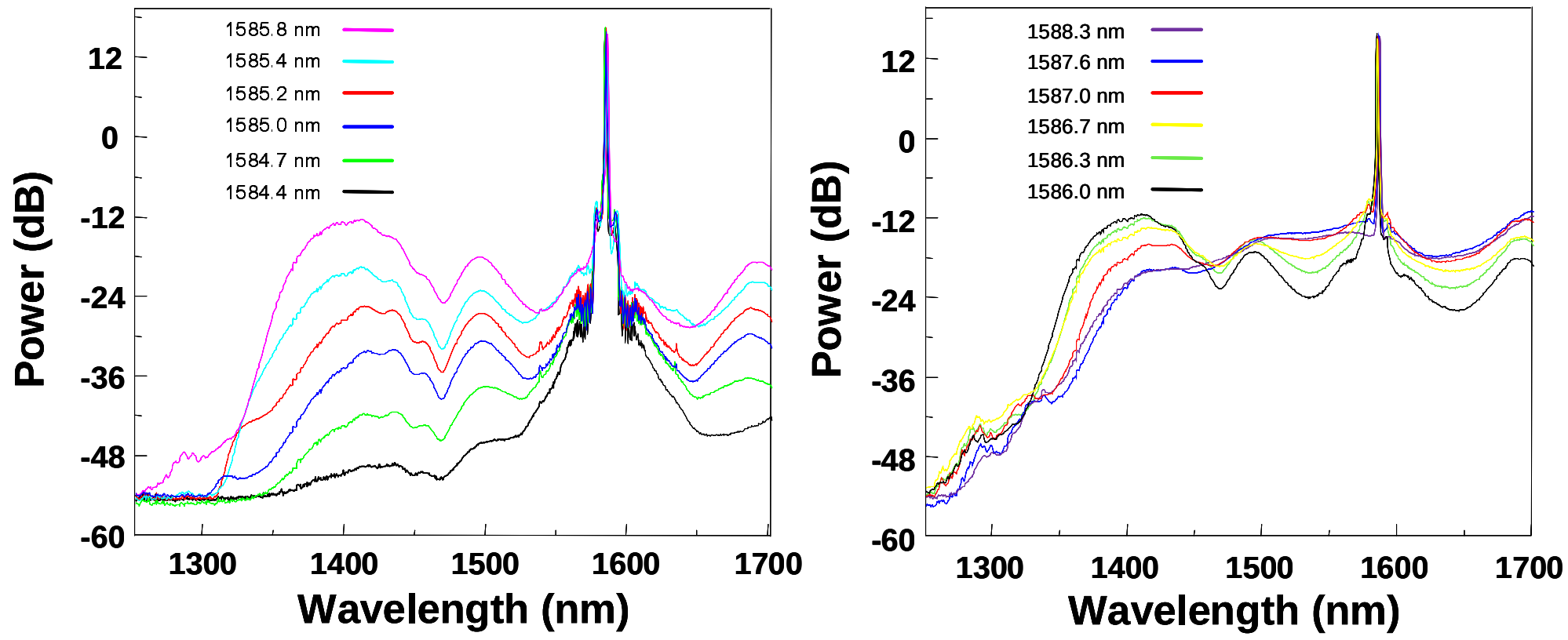

Fig. 9. OPA gain spectra as a function of pump wavelength. (a) $\lambda_p$ tuned from 1584.4 nm (black line) to 1585.8 nm (magenta). (b) $\lambda_p$ tuned from 1586 nm (black line) to 1588.8 nm (gray).

The measurements in Figs. 8 and 9 were taken with a duty cycle of 1/4000 and an OSA resolution of 1-nm. It was observed that the amount of gain experienced had a greater dependence on the pump spectral filtering and less on the duty cycle. Thus, it is expected that an improvement in noise filtering around the pump is a viable means to increase the gain bandwidth. In the same direction, improving the high pump power generation should reduce the level of input noise thus allowing a bandwidth extension. Other means to avoid this noise mediated gain compression is using HNLFs with negative $\beta_4$ and pumping in the normal region to produce narrowband regions of gain as shown in [20].

Finally, note that some applications in the SWIR band require high output power. Optical parametric oscillators are recognized as good candidates to fulfil these requirements, since a good fraction of the pump power can be transferred to the signal and the idler [21,22]. In the reported scheme, it was verified that simply by adjusting the input signal power to appropriate levels, it is possible to produce an output signal power of up to 15 W.

## 4. Conclusion

In conclusion, by employing an HNLF having low and positive fourth order dispersion coefficient, we have demonstrated a considerable parametric gain and conversion over 730 nm between the NIR (1300 nm) and SWIR (2030 nm) bands using a silica-fiber device. The peak gain was 69.5 dB at 1820 nm, with an average gain of 40 dB in the 2000 nm band. Furthermore, from noise gain measurements we estimated that the OPA bandwidth can reach 840 nm and further improvements in generating high pump powers might bring this value to 1000 nm or more. Our results demonstrate that the highly-developed NIR-based infrastructure can be leveraged for application in the SWIR band by employing a parametric conversion process in a highly nonlinear silica fiber.

## Acknowledgments

This work was supported in part by the Lockheed Martin Corporation and by the Defence Advanced Research Project Agency. We thank S. Moro for providing the 1390 nm laser.